\documentclass[twocolumn,preprintnumbers,amssymb,amsmath,9pt]{revtex4}[10pt]
\usepackage{graphicx}% Include figure files
\usepackage{dcolumn}% Align table columns on decimal point
\usepackage{bm}% bold math
\usepackage{amssymb}
\usepackage{epsfig}
\newcommand{\be}{\begin{equation}}
\newcommand{\ee}{\end{equation}}
\newcommand\beq{\begin{eqnarray}} 
\newcommand\eeq{\end{eqnarray}}

\newcommand{\GeV}{{\rm ~GeV }}

\newcommand{\phiback}{\phi^{(0)}}
\newcommand{\lya}{{Ly \alpha}}

\begin{document}

\title{Astrophysical Effects of Scalar Dark Matter Miniclusters}
\author{Kathryn M. Zurek, Craig J. Hogan, Thomas R. Quinn}
\affiliation{Physics and Astronomy Departments, University of Washington, Seattle, WA
98195}

\preprint{}
\begin{abstract}

We model  the formation, evolution and astrophysical effects of dark compact  Scalar  Miniclusters (``ScaMs''). These objects arise when a
scalar field, with an axion-like or Higgs-like potential, undergoes a
second order phase transition below the QCD scale. Such a scalar 
  field may couple too weakly to the standard
model to be detectable directly through particle interactions, but may still be
detectable by gravitational effects,   such as  lensing and baryon accretion by  large, gravitationally bound miniclusters.   The masses of
these objects are shown to be constrained by the Ly$\alpha$  power spectrum to be less
than $\sim 10^4 M_\odot$, but they may be as light as classical axion miniclusters, of the order of
$10^{-12} M_\odot$.   
 We simulate the formation and nonlinear gravitational collapse of these objects around matter-radiation equality using an N-body code,   estimate their gravitational lensing properties, and assess the  
  feasibility of studying   them using current 
  and future lensing experiments. Future MACHO-type variability surveys of many background sources can reveal either high-amplification, strong lensing events, or  measure density profiles directly via weak-lensing variability, depending on ScaM parameters and survey depth.  However, ScaMs, due to their low internal densities, are unlikely to be responsible for apparent MACHO events already detected in the Galactic halo.  As a result, in the entire window  between $10^-7 M_\odot$  and $10^2 M_\odot$ covered by the galactic scale lensing experiments, ScaMs may in fact compose all the dark matter. 
A simple estimate is  made of parameters that would give  rise to  early structure formation; in principle, early stellar collapse could be triggered by ScaMs as early as recombination, and significantly affect cosmic reionization.

%which was motivated in ref.~\cite{Kaplan:2005wd}, where it was shown that the late phase transition can also give rise to interesting signals in axion detection experiments.
%We consider a scalar field theory with phase transition around the QCD time which yields compact objects of mass $M \sim 1 M_\odot$ and densities which make them good candidates for gravitational lensing experiments SuperMACHO and POINT-AGAPE.  These objects arise in theories such as that proposed in ref.~\cite{Kaplan:2005wd}, where it was shown that the late phase transition can also give rise to interesting signals in axion detection experiments.
\end{abstract}

\maketitle

\section{Introduction}

Despite experimental efforts to ascertain  the nature of dark matter, its fundamental character
 remains a mystery.  Particles motivated by
supersymmetry (Weakly Interacting Massive Particles (WIMPs)), extra
dimensions (Kaluza Klein Particles), and a solution to the strong CP
problem (the pseudoscalar axion), are among the best motivated from a
particle physics standpoint (see \cite{Bertone:2004pz,PDBook} for reviews).  Experimental efforts searching for such elementary particle
dark matter have focused on utilizing their strong or electroweak
interactions to detect them, either directly through their
interactions with ordinary matter, or indirectly through their
annihilations to photons.  

%Scalar dark matter candidates in these theories abound, but are almost always associated with physics above the electroweak breaking scale (even the axion, which is light, is associated with physics at the Peccei-Quinn scale, around $10^{10}\GeV$).  In other realms of cosmology, such as theories of inflation and dark energy, scalars are also in rampant use; in the case of dark energy, the scalars are also typically extremely light (sub-eV or lighter).

%he DAMA experiment has reported a tantalizing signal consistent with a dark matter particle with mass in the range, , while the CDMS experiment rules out almost completely a particle in this range (although see \cite{resolution} for possible explanations of this apparent contradiction).  Experiments are also under way searching for axions.  No dark matter signal has yet been detected from the axion, however, the sensitivity of these experiments is reaching a point where it is expected that the axion will either be detected or ruled out as a dark matter candidate in the next generation of experiments \cite{Rosenberg:2004zc}.   
On another front, gravitational lensing has already proven an effective way to  probe the  nature of dark matter  experimentally. Photometric monitoring of many stars has been used to search for gravitational lensing
by lumps of dark matter, generically known as Massive Compact Halo Objects
(MACHOs)\cite{Alcock:2000ph,Udalski:2002jg,Aubourg:1993wb,Tisserand:2006zx}.
%In addition to searching for planetary baryonic dark matter
%(which is now ruled out), it was also hoped that such experiments
%would be able to detect black hole dark matter, as some  models (for example holes produced during the QCD phase
%transition as in \cite{Jedamzik:1996mr}) yield black holes with masses
%in the solar mass range, where lensing experiments were most sensitive.  These are now ruled out, 
Dark matter dominated by compact objects such as baryonic planets, stellar remnants or black holes  is now only allowed for  $m_H \lesssim 10^{-7} M_\odot$ and $m_H \gtrsim 30 M_\odot$ \cite{Bennett:2001vh}.
However, there remains the ``microlensing puzzle'', which is that MACHOs are observed  and appear to   contribute an optical depth toward the Large Magellenic
Cloud (LMC)   too large by a factor $\sim 5$ to be accounted for by
simple models of the stellar population.  Although the results are not
consistent with all of the dark matter being in the form of MACHOs,
their result \cite{Alcock:2000ph} is
consistent with an object of mass $M \sim 0.4 M_\odot$ accounting
for $20\% $ of the halo mass.  This signal is also at variance with
the EROS experiment results \cite{Tisserand:2006zx}, which exclude $0.4 M_\odot$ MACHOs from composing more
than 7\% of the halo.

%This result and its interpretation has been controversial (see for example
%\cite{Evans:2004gd,Belokurov:2004am,Green:2002qk,Jetzer:2002ur}), and
%it has not been clear whether the events are MACHOs in the halo or are
%the result of self lensing by baryonic stars or remnants close to the LMC.  The POINT-AGAPGE experiment
%\cite{CalchiNovati:2005cd}, searching for microlensing in the
%direction of M31, also reports a lower bound for objects $M \sim 0.5-1
%M_\odot$ of at least 20\% of the halo mass, although the lower bound
%drops to 8\% for objects with masses $M \sim 10^{-2} M_\odot$.  %One
%possible explanation for the MACHO and POINT-AGAPE data is that their
%signal is the result of lensing of stellar remnants like neutron stars
%or brown dwarfs.  

Searches for MACHO-like objects in the halo continue
\cite{Rest:2005jn,Tisserand:2006zx}, and in the future,  lensing searches for objects on cosmological
scales will be increasingly sensitive to objects in a wider range
of masses (see \cite{2005ApJ...618..403B,2003ApJ...589..844W} for
proposals on cosmological scale lensing using gamma-ray bursts, and
\cite{1994ApJ...424..550D,2002ApJ...577...57W,2003PhDT.........7W,1996MNRAS.278..787H} for gravitational
lensing from distant quasars).  In addition, a large-aperture, wide-field spaceborne telescope is   potentially   capable of
monitoring individual stars for lensing effects in a program similar to MACHO, but  in   galaxy halos   orders of magnitude farther away than the current
galactic lensing experiments (see \cite{Baltz:2003ds} for lensing
observations towards M87 in the Virgo cluster using the Hubble Telescope). In more distant halos,  the Einstein
radius for lensing by a given mass is considerably larger;
therefore a wider range of objects can produce observable
microlensing, and the probability of events increases.

The continued increasing capabilities in lensing
experiments
opens the possibility of detecting lensing variability  from a wider class of astrophysical objects, beyond baryonic planets, stellar remnants and black holes, and has the potential for studying certain types  of non-baryonic dark matter in detail.  In particular, in this paper we study the effects of a type of  non-baryonic,
elementary particle dark matter that naturally forms into large,
self-gravitating clumps in the early universe:  scalar dark matter
miniclusters. These structures originate from  order unity isocurvature matter  density fluctuations created during a second order phase
transition of a very light scalar field, sometime after the QCD phase
transition.   The miniclusters   form  when these fluctuations subsequently collapse gravitationally at temperatures near matter-radiation equality.

It has been known for some time that the QCD phase transition gives
rise to dense axion configurations, originally called miniclusters \cite{Hogan:1988mp},
or alternatively axitons \cite{Kolb:1993zz,Kolb:1994fi}, with masses $M_{mc} \sim
10^{-12} M_\odot$, which are detectable by pico- or femto- lensing
experiments \cite{Kolb:1995bu}.  The effects of large-scale modulation of axion density was also studied in \cite{Khlopov:1999tm,Sakharov:1996xg,Sakharov:1994id}.  We consider here a similar mechanism for a much wider
class of theories, arising from a second order phase transition in either a Higgs-like
or axion-like system, and with an associated phase transition
temperature possibly much lower, perhaps even well below the QCD
temperature. For this class of theories, the gravitational effects of the miniclusters are the main distinguishing effects of the character of the dark matter and the main experimental constraint on their parameters.

These scalar miniclusters (which we designate ScaMs for short)  may be in a mass range interesting for
lensing experiments.  Although their densities are typically too low to be detectable by current generation galactic lensing experiments (and as a result, the MACHO dark matter constraints do not apply), they may be seen through longer baseline galactic lensing or cosmological scale lensing.  As shown below, masses as large as $\sim10^4 M_\odot$ are currently allowed without contradicting constraints on the power spectrum from   the Ly$\alpha$ forest data.  Other constraints on MACHO-like objects  from tidal effects in halo wide binaries (as in \cite{Yoo:2003fr}) also do not apply to ScaMs on account of the ScaMs' low internal densities. 
A significant fraction (as much as half) of the dark matter collapses into the compact objects
initially, so it is natural to find significant microlensing rates under  the right conditions.  Unlike true MACHOs however, these objects are not point-like
gravitational sources, but are extended objects with a radius which can be
on the same order or larger than the Einstein radius, depending on the distance to the lens.  This leads to the
possibility of unique and distinctive gravitational lensing signatures. In the strong lensing regime, the ScaMs produce classical caustic events with sudden appearance and disappearance of images, associated with sudden large-amplitude variations in image brightness in variability surveys. In the weak lensing regime, sources passing behind a ScaM experience variable small-amplitude modulation depending on the projected surface density of the dark matter. 

There are a wide range of theories which generate such objects.  The
second-order scalar field phase transition was introduced in
ref.~\cite{Kaplan:2005wd} within the context of axion cosmology to
show that in the presence of late phase transitions the axion mass and
coupling constant may lie outside the window prescribed by the
conventional astrophysical and cosmological constraints;  in particular
a string scale axion with decay constant $f = 10^{16}\GeV$ is
allowed. In this model with a Higgs-like potential with a very small
mass term, the field remains in the unbroken phase until the curvature
of the potential is sufficiently large to overcome the Hubble
friction, at which time the symmetry breaks as the field evolves to the minimum
of the potential.  Pseudo-Goldstone bosons have also been invoked for a
variety of uses, the most famous of which is the axion to solve the
strong CP problem.  They have been used in relation to attempts solve
the cosmological constant problem \cite{Frieman:1995pm}, explain the origin of large-scale
structure \cite{Press:1989id,Frieman:1991tu}, and provide a warm dark matter
candidate \cite{Hogan:1994sa}.  The considerable
increase in our knowledge of cosmological parameters has ruled out or
disfavored a number of these scenarios, however the
pseudo-goldstone boson remains a viable dark matter candidate. 

The microlensing experiments have the capability of detecting these
scalar objects which may never be observed directly through particle
interactions due to their very weak couplings.  The weak couplings are
required by naturalness arguments.   If there is no symmetry to protect their
masses, radiative corrections tend to force scalars to be as heavy as
the cut-off scale, which, for electroweak SUSY breaking, is
$m_{susy} \sim 100 \mbox{ GeV}$.  In order to maintain a
scalar as light as $\Lambda_{QCD}^2/M_{pl}\sim 10^{-21} \mbox{ GeV}$, this scalar must be
protected from this SUSY breaking; this is done by requiring a small
enough coupling to the visible sector that sufficiently small
radiative corrections are generated for the light scalar field.  We
require in particular that its coupling $\lambda$ to all ordinary
matter satisfy $\lambda
< m_\phi/m_{susy}$.   Such an object is truly dark, undetectable even by the most sensitive particle dark matter detectors; like black holes, we can detect their presence only through their gravitational interactions. 

In addition to lensing, these objects seed non-standard bottom-up hierarchical
structure formation.  The phase transition injects a large amount of fluctuation
power on small scales so that nonlinear clusters are predicted already
at recombination; they can accrete baryons, and potentially trigger
star formation (possibly assisting early re-ionization), much earlier than the standard dark matter with only
inflationary perturbations.  
 This also implies that measurements of the matter power spectrum from the Lyman-Alpha forest will limit the masses of these objects; we   derive constraints below.

The plan of the paper is as follows. In the next section we
describe the physical dark matter models of interest, and derive general expressions for the
density fluctuations.  We then discuss the phenomenology of
ScaMs, including the limits on their masses and radii from
microlensing experiments and measurements of the power spectrum of the
Lyman-Alpha forest.   We use an N-body code to determine their density profiles,
and consider their   gravitational lensing effects in two regimes, strong and weak lensing, corresponding to distant and nearby halos respectively.  We briefly discuss the evolution of these objects and
their impact on structure formation, in particular early star
formation.  We close by surveying the   viability of detecting and studying this
dark matter candidate through its lensing effects.

\section{ScaMs and a late cosmological phase transition}

\subsection{Formation of isocurvature fluctuations}

Consider the generic potential of a complex scalar field $\phi=\rho e^{i a/f}$,
\be
V(\rho,a) = -m_\rho^2 \rho^2 +\lambda \rho^4 + \mu^4 \left(1-\cos\frac{a}{f}\right)
\ee
%V=\mu^4\left|\frac{\phi^2}{f^2}-1\right|^2;
%\ee
%such a potential could be generated from a supersymmetric theory to protect the mass of the scalar.  If we rewrite $\phi = \rho e^{i a/M}$, we have
%\be
%V=\mu^4\left(\frac{\rho^4}{f^4}+1-\frac{\rho^2}{f^2}\cos\frac{a}{f}\right).
%\ee
Such a potential is similar to that generated for the QCD axion, the
difference being that QCD instanton effects result in $\mu$ effectively being a
time dependent quantity.  That such a potential generates ${\cal
  O}(1)$ fluctuations in the resulting dark matter has been known for
some time \cite{Hogan:1988mp}.  We review the basics here.  If the
Peccei-Quinn symmetry breaks after inflation has already occurred, it
is expected that the initial value of the field will vary spatially,
$-\pi \leq \theta \leq \pi$.  As the QCD phase transition nears at $T
\simeq 1 \mbox{ GeV}$ the axion gains a mass due to QCD instanton
effects, generating a potential for the axion favoring $\theta = 0$.
Initial spatial fluctuations in the field, $\theta_i$, are then
translated into spatial fluctuations in dark matter density,
$\rho_{DM} \propto \theta_i^2$.  As the fluctuations in $\theta_i$ are
${\cal O}(1)$, the axion dark matter fluctuations are also expected to
be ${\cal O}(1)$.  Thus the matter power spectrum enters
matter-radiation equality already non-linear.  These
fluctuations immediately collapse at matter-radiation equality into
dense objects, ScaMs.  For the particular case of axions, the
mass of these objects was shown to be approximately $10^{-12}M_\odot \sim
\rho_a(T\simeq1\mbox{ GeV}) d_H^3$, the total dark matter mass within the horizon, $d_H$, at
the time of the QCD phase transition.  

These
arguments carry over generally to any type of pseudo-Goldstone boson,
provided, like the axion, the symmetry for the radial mode breaks after inflation, and
the angular mode is effectively massless when this symmetry breaking
occurs.  Like the axion, a phase transition generates large density
fluctuations, but we allow the transition to occur potentially much
later, giving rise to more massive ScaMs.  In the appendix we
give an example of a specific supersymmetric model which gives rise to
such late phase transitions.  We call these pseudo-Goldstone boson modes axion-like modes on account of the similarity to the axion itself.  These axion-like modes, however, need not be connected to QCD physics in any way.

In similar fashion, we may also consider Higgs-like modes.  If the radial mode is sufficiently light, a Higgs-like potential may also generate a late second-order phase transition when $\rho$ breaks the $U(1)$ symmetry. We will see that this second-order  phase transition also results in ${\cal O}(1)$ density fluctuations in dark matter condensate.   The radial field $\rho$ remains in an unbroken phase at the origin, $\rho_i=0$, until the curvature there exceeds the Hubble friction, $m_\rho > H$, when the field rolls out to its true minimum at $\langle\rho\rangle = f$.  We will show that quantum fluctuations about $\rho_i$ result in ${\cal O}(1)$ variations in the the roll-off time of the scalar field.  These ${\cal O}(1)$ variations in the roll-off time translate to ${\cal O}(1)$ variations in the density, due to spatial modulations in  the temperature when the dark matter condensate forms and begins to redshift.  In particular, we calculate
\be
\frac{\delta\rho}{\bar{\rho}}=\frac{\bar{T}_{trans}^3}{T_{trans}^3}-1,
\ee
where bars denote mean values, $T_{trans}$ is the phase transition temperature in a particular Hubble patch, and
\be
\bar{\rho}(T) = m_\rho^2 f^2 \left(\frac{T}{\bar{T}_{trans}}\right)^3.
\ee
It will be useful to rewrite this in terms of a time delay, $\delta t$ from the mean time of the phase transition, $\bar{t}_{trans}$,
\be
\frac{\delta\rho({\bf x})}{\bar{\rho}}=\left(\frac{t_{trans}({\bf x})}{\bar{t}_{trans}}\right)^{3/2}-1=\left(1+\frac{\delta t({\bf x})}{\bar{t}}\right)^{3/2}-1,
\ee
where we have included the explicit spatial dependence of $\delta\rho({\bf x})$ on scales exceeding the Hubble size at the time of the phase transition.
We can see that in the limit that $\delta t \lesssim \bar{t}$, the result reduces to
\be
\frac{\delta\rho({\bf x})}{\bar{\rho}}\simeq \frac{9}{2}H(\bar{T}_{trans})\delta t({\bf x}),
\ee
which is identical to the inflationary result for an upside-down harmonic oscillator \cite{Guth:1985ya}, up to the multiplying factor.

As a result, in order to calculate the density fluctuations, we need only calculate the time delay.   
This is approximated by \cite{Guth:1985ya}
\be
\delta t({\bf x}) \simeq \lim_{t\rightarrow\infty}\frac{\delta \phi({\bf x},t)}{\dot{\phi}^{(0)}({\bf x},t)},
\ee
where we have expanded in fluctuations, $\delta\phi({\bf x},t)$, around the background field, $\phiback$, $\phi({\bf x},t) = \phiback(t) + \delta\phi({\bf x},t)$. 

We determine the classical evolution
of the background field $\phiback$ from its equation of motion,
\be
\ddot{\phi}^{(0)}(t)+3H\dot{\phi}^{(0)}(t)+m^2\phiback(t)=0.
\ee
%whose solution, assuming a rapid phase transition so that $H$ is approximately constant, is given by 
%\be
%\phiback(t) = \phiback_0(t=0) e^{H t F(m^2/H^2)},
%\ee
%where
%\be
%F(m^2/H^2) = \sqrt{\frac{9}{4}+\frac{m^2}{H^2}}-\frac{3}{2}.
%\ee
Since we are interested in the behavior at asymptotically late times, we ignore the Hubble friction terms, and the solution is
\be
\phiback(t) = \phiback_0(t=0) e^{m t}.
\ee
%Taking the transition to occur when $3 H \simeq m$, we then find its velocity as it begins to roll is
%\be
%\dot{\phi}^{(0)} \simeq \phiback_0 H F(m^2/H^2){H t F(m^2/H^2)}.
%\ee
Likewise, we can determine $\delta\phi({\bf x},t)$ from an equation of motion
\be
\ddot{\delta\phi}({\bf x},t)+3 H \dot{\delta \phi}({\bf x},t)+m^2\delta\phi({\bf x},t)-\nabla^2\delta\phi({\bf x},t)=0.
\ee
For large times, we may neglect the spatial gradient and Hubble friction terms, and the solution is
\be
\delta\phi({\bf x},t)=\delta\phi({\bf x},t=0)e^{m t}.
\label{deltaphi(x)}
\ee
%\be
%\delta\phi({\bf x},t)=\delta\phi({\bf x},t=0)e^{H t F(m^2/H^2)}.
%\label{deltaphi(x)}
%\ee
Thus $\delta t({\bf x})$ is set entirely by initial conditions:
\be
\delta t({\bf x})\simeq \frac{\delta\phi({\bf x},t=0)}{\phi^{(0)}(t=0)}\frac{1}{m}.
\ee

We will assume that the background field $\phiback$ is initially at the origin. However, we cannot choose $\phiback(t=0)=0$ because of quantum fluctuations.  The size of these fluctuations are calculated from the Fourier transform of the two point function, following \cite{Guth:1985ya}:
\begin{eqnarray}
\Delta \phi ({\bf k})& \equiv & \left[\frac{k^3}{(2\pi)^3}\int \frac{d^3x}{(2\pi)^3} e^{i{\bf k\cdot x}}\langle\delta\phi({\bf x},t=0)\delta\phi(0)\rangle\right]^{1/2} \nonumber \\
&=& \left[ k^3\int \frac{d^3k'}{(2\pi)^3}\int \frac{d^3x}{(2\pi)^3} e^{i{\bf k\cdot x}} e^{i{\bf k'\cdot x}}  \right]^{1/2} \nonumber \\
&=& \frac{k}{(32\pi^3)^{1/2}}
\label{rmsDeltaphi}
\end{eqnarray}
For the mode of interest, $k = m$, these fluctuations set both $\phiback(t=0)$ and the root mean square (rms) value of $\delta\phi({\bf k})$, the Fourier transform of $\delta\phi({\bf x},t=0)$.
Putting it all together, we have the rms density fluctuations,
\beq
\frac{\Delta \rho_{rms}}{\bar{\rho}} &\simeq& \frac{9}{2}\frac{\Delta\phi(k=m)}{\phiback(t=0)}\frac{H(\bar{T}_{trans})}{m}\nonumber \\
&\simeq& \frac{1}{2\sqrt{2}},
\label{Higgs-like-flucs}
\eeq
showing that the density fluctuations are of order unity.

\subsection{ScaM collapse}
These fluctuations   collapse gravitationally into ScaMs around the time of matter-radiation equality. For collapse of a uniform sphere the final core density of a virialized
 ScaM is \cite{Kolb:1994fi}
\be
\rho = 140 \delta^3(\delta+1) \rho_{eq},
\label{isothermal-density}
\ee
where $\delta = \delta \rho/\bar{\rho}$.  Their masses are set by the dark matter mass inside the horizon at the time of the phase transition,
\be
M_{ScaM} = \frac{4}{3}\pi d_H^3 \rho_{DM}(T_{trans}),
\ee
where $d_H = H(T_{trans})^{-1}$.  This corresponds to a ScaM mass, assuming these scalars compose a fraction $r$ of the dark matter,
\be
M_{ScaM}\simeq 1.4\times 10^{-3} M_\odot r\left(\frac{10^{-3}\mbox{ GeV}}{T_{trans}}\right)^3.
\label{TempMassRel}
\ee

\subsection{Limits on ScaM Mass from Lyman-$\alpha$}

As a result of these non-linear density fluctuations, the phase
transition adds a large amount of fluctuation power into the  spectrum
on small scales.  These scales, $r_s \sim T_{trans}/(T_0
H(T_{trans}))$  ($\sim10\mbox{ pc}$ (comoving) for $T_{trans}\sim \Lambda_{QCD}$)
are well below the reach of current measurements of the power
spectrum, since the smallest scale measurements,   derived from
observations of the Lyman-$\alpha$ absorption of the spectra of
distant quasars \cite{2005ApJ...635..761M,2006ApJS..163...80M,2000ApJ...543....1M}, reach down only to scales $\sim 0.1 h^{-1}\mbox{
  Mpc}$, the scale where protogalactic gas is collapsing into mildly
nonlinear filaments.   The phase transition, however, generates a Poisson white noise power spectrum on scales larger than $r_s$ which adds to the inflationary power on  the Lyman-$\alpha$ scale,  
\be
P_p = \frac{1}{n_{ScaM}},
\label{power-spectrum}
\ee 
where $n_{ScaM}$ is the number density of ScaMs and the subscript $p$
is for primordial.  The added power today is then the product of the
primordial white noise power spectrum with the transfer function for isocurvature fluctuations, $T_{iso}$,
\be
P_{wn} = T_{iso}^2 P_p,
\ee
where
\be
T_{iso} = \frac{3}{2}(1+z_{eq}).
\ee

Now neither $P_p$ nor $T_{iso}$ is wave number $k$ dependent, whereas
$P_{CDM}$ decreases with $k$.  We plot this power spectrum in
fig.~1.  We have introduced in the white noise spectrum a
smoothing scale $r_s \simeq d_H(T_{trans})$, on which the Kibble mechanism
smooths field fluctuations, to remove power on the smallest scales,
\be
P_{ScaM} = P_{wn} e^{-(k r_s/2\pi)^2/2} .
\label{Pscam}
\ee  

We can see from fig. 1 that for sufficiently large $k$, $P_{ScaM}$ will exceed $P_{CDM}$; we must ensure that this occurs on smaller scales than are reachable with Lyman-alpha measurements, $k>k_J$, so that we require 
\be
\frac{9}{4}(1+z_{eq})^2 \frac{M_{ScaM}}{\rho_{DM}} < P_{\lya}(k_J),
\ee
where $k_J\simeq 10 h\mbox{ Mpc}^{-1}$.  This yields the constraint
\be
M_{ScaM} \lesssim 4 \times 10^3 M_\odot,
\ee
corresponding to a constraint on the temperature of the phase transition, 
\be
2 \times 10^{-5} \mbox{ GeV} \lesssim T_{trans},
\ee
assuming the scalars compose all the dark matter.  A similar
constraint from the Lyman-alpha power spectrum was derived in
\cite{Afshordi:2003zb} using numerical simulations in the context of
primordial black holes.  

\begin{figure}
\includegraphics[width=8.5cm]{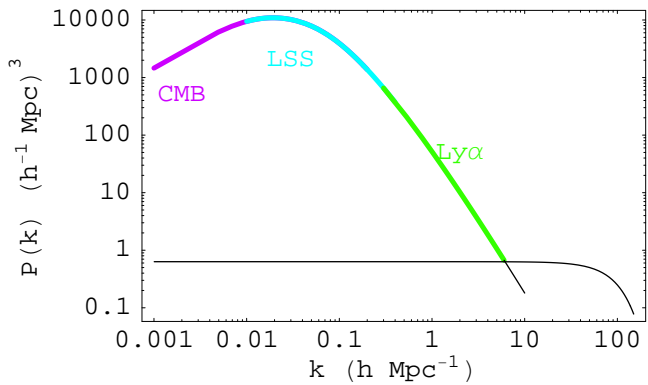}
\caption{CDM power spectrum derived from BBKS \cite{1986ApJ...304...15B} with white
  noise power spectrum smoothed on scales $r_s \simeq 7 \times 10^{-2}
  \mbox{ Mpc}$, given by eqn.~\ref{Pscam}.  The amplitude of the white noise spectrum corresponds
  to the power spectrum for a ScaM of mass $ M \simeq 4 \times
  10^3 M_\odot$.}
\label{white-noise-power}
\end{figure}

Phase transitions between the QCD scale and this Lyman-$\alpha$ limit create abundant ScaMs  in the mass range
\be
10^{-12} M_\odot \lesssim M_{ScaM} \lesssim 4 \times 10^3 M_\odot,
\ee
much of which is accessible to current and future microlensing experiments.  

\subsection{ScaMs as microlenses}

For ScaMs to act as observable strong microlenses, three conditions must be satisfied:

\begin{enumerate}

\item ScaM masses must lie in the mass range reachable by experiments.
  For classic microlens searches by stellar monitoring in the local
  halo, this range is currently $10^{-7}M_\odot \lesssim M_{ScaM}
  \lesssim 10 M_\odot$.  However, the accessible range will widen in the future as microlensing experiments access fainter and more distant monitored background sources in other galaxies.

\item  A rough criterion for  strong lensing, leading potentially to
  large-amplitude variations in source brightness, is that the radius of the ScaM be smaller than the Einstein ring radius.  For microlensing of objects over cosmological distances, the Einstein radius is
\be
R_E \simeq 3 \times 10^{16} \left(\frac{M}{1 M_\odot}\right)^{1/2}\mbox{ cm}.
\ee
For lensing toward a source in the local galactic neighborhood (e.g. toward the LMC or M31), the Einstein Radius is
\be
R_E \simeq 3 \times 10^{14} \left(\frac{M}{1 M_\odot}\right)^{1/2}\left(\frac{D}{50\mbox{ kpc}}\right)^{1/2}\mbox{ cm},
\ee
where $D$ is the distance to the lens, and it is assumed that $D \ll D_s$, the distance to the source.

Using the spherical model, eqn.~\ref{isothermal-density}, and assuming that the ScaMs are approximately
constant density, we calculate
\be
R_{ScaM} =  4 \times 10^{16} \frac{1}{\delta\left((\delta+1)\Omega_\phi\right)^{1/3}}\left(\frac{M_{ScaM}}{1 M_\odot}\right)^{1/3}\mbox{ cm}.
\label{R}
\ee

\item Their cosmological abundance must be consistent with the limits
  from the current lensing experiments.  Since these objects would
  generally be too fluffy to create strong lensing in the nearby halos observed by the current generation of
  galactic microlensing experiments,  consistency with the
  limits of these experiments is generally not problematic.

\end{enumerate}
Although these objects would generally not be dense enough to be observed by Galactic microlensing experiments, they may be detectable as microlenses for more distant sources and halos. If they do produce strong lensing events, they do not obey the classic Paczynski \cite{Paczynski:1985jf} point-mass light curve, but instead are dominated by more generic caustic-crossing events.  More generally, in nearby halos they may not even act as strong lenses, but may have a resolved density structure that appears as small-amplitude variations in the light curve of a lensed source.  

  To improve on  the spherical collapse model, eqn.~\ref{isothermal-density}, and
  in particular  to determine properties of these objects observable
  by lensing experiments, we simulate the collapse of ScaMs using an N-body code. The resulting objects are more realistic than the spherical model and allow  determination of  some representative density profiles.

%In making the mass temperature correspondance here we have assumed the scalar fraction of the dark matter is 50\%.  This can occur without any fine tuning, as shown in the appendix, when the axion field delays the time of the phase transition.  In addition, as we found section II, we expect $\sim 50\%$ of the scalar dark matter to collapse into ScaMs.  Then we have $\sim25\%$ of the dark matter in scalar ScaMs.  If the axion field does not delay the time of the phase transition, we would expect closer to $\sim50\%$ of the dark matter in these compact objects.  

%This is an interesting mass range from the perspective of microlensing.  On the lower end of this mass range, the microlensing experiments rule out a significant fraction of the dark matter halo in objects with $M_{ScaM} \gtrsim 10^{-7} M_\odot$.  At masses $M_{ScaM} \sim 0.1-1 M_\odot$ the experiments are consistent with $\sim 20 \%$ of the dark matter halo in such objects, with the limit loosening for more massive objects, and becoming totally unrestrictive for $M_{ScaM} > 30 M_\odot$ \cite{Bennett:2001vh}.  A phase transition right at the QCD scale, leading to an object with mass, $M_{ScaM} \sim 7 \times 10^{-9}M_\odot$, is also unrestricted by the microlensing data. 

%We have not yet considered subsequent evolution of the ScaMs after their initial formation at equality.  In particular we wish to make some analytic estimates of how the ScaMs dissipate or accrete matter.

\section{Simulating ScaMs}

 We simulate the
 formation of ScaMs in the radiation dominated era using the N-body code described in \cite{2004NewA....9..137W}.

\subsection{Initial conditions} 

The initial density profile may be determined utilizing one of two
methods: either by solving the classical equations of motion for a
field $\phi$ directly, or simply using the power spectrum of
eqn.~\ref{power-spectrum}.   The equation of motion for a scalar field
is given simply by  
\be
\ddot{\phi}+3H\dot{\phi}-\frac{1}{R^2(t)}\nabla^2\phi+\frac{\partial V(\phi)}{\partial \phi}=0,
\ee
where the Laplacian is taken with respect to comoving coordinates $x$.
We can rewrite this (see \cite{Kolb:1994fi} for details) in terms of conformal time, $\eta\equiv R/R_1$, where $R_1$ is defined by $H(R_1)=m_\phi$, and comoving Laplacian $\bar{\nabla}^2$, taken with respect to coordinates $\bar{x} = H(R_1) R_1 x$,
\be
\phi''+\frac{2}{\eta}\phi'-\bar{\nabla}^2\phi+\frac{\eta^2}{m_\phi^2}\frac{\partial
  V(\phi)}{\partial \phi}=0.
\label{diff-eq}
\ee
We assume the system is subject to white noise initial conditions,
\beq
\phi_i=A\sum&&\frac{\sin(\omega \eta)}{\omega \eta}\sin(p_i x+\xi_{1ijk})\sin(p_j y+\xi_{2ijk})\nonumber\\
&&\times\sin(p_k z+\xi_{3ijk}),
\eeq
where the $\xi$'s are random phases.

We solve the equation of motion on a lattice 100 sites per side for the
axion-like potential, $V(\phi) = 1-\cos(\phi/f)=1-\cos\theta$.  This potential
would result in the formation of domain walls if, for example,
$A=\pi f$ (in the language of axions, this corresponds to multiple vacua, $N > 1$).  The walls are transient objects that quickly dissipate by particle radiation, but introduce singularities that make numerical integration difficult. Since we are interested in the formation and evolution of
the ScaMs only, we avoid this problem by choosing $A$ so that $\theta$ varies between $-\pi$ and $\pi$, and the root-mean square (rms) value of $\theta_i = \phi_i/f$ is the rms average value of the misalingment angle,
$\theta_{rms}=\pi/\sqrt{3}$.  In this case, the domain walls never form in our box.    

%In the case of the $\phi^4$ potential, these can be removed by a small $\phi^3$ term breaking the $Z_2$ symmetry.  As we are interested here in the evolution of the ScaMs and not the domain walls, we simply avoid this problem by offsetting $\phi_i$ slightly from zero.  Likewise, with the axion potential, we avoid the formation of domain walls by choosing $A=\pi$, a sufficiently small amplitude.  We discuss later possible implications of choosing a larger (more physical) amplitude.

\begin{figure}
\includegraphics[width=8.5cm]{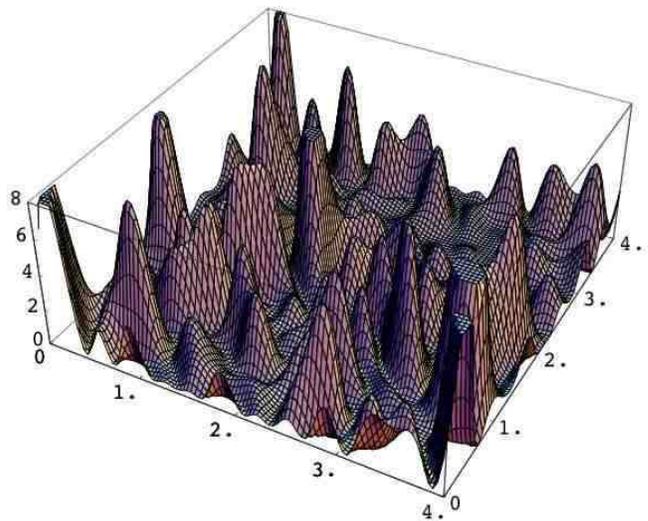}
\caption{Two-dimensional slice in the x-y plane of initial white noise
  field energy density distribution.  x-y coordinates are in $\eta$, where
  $\eta = 1$ corresponds to a length $d_H(T_{trans})$, one horizon size at the time of the phase transition.  The z-axis is the initial white noise over-density $\rho({\bf x})/\bar{\rho}$, where $\bar{\rho}$ is the mean density in the box.}
\label{white-noise}
\end{figure}

Solving the equation of motion numerically automatically simulates the
effects of the Kibble mechanism, smoothing field fluctuations on
scales smaller than the horizon size at the time of the phase
transition.  We plot in fig.~\ref{white-noise} a two dimensional slice of the
initial white noise density fluctuations for the axion-like potential.

\begin{figure}
\includegraphics[width=8.5cm]{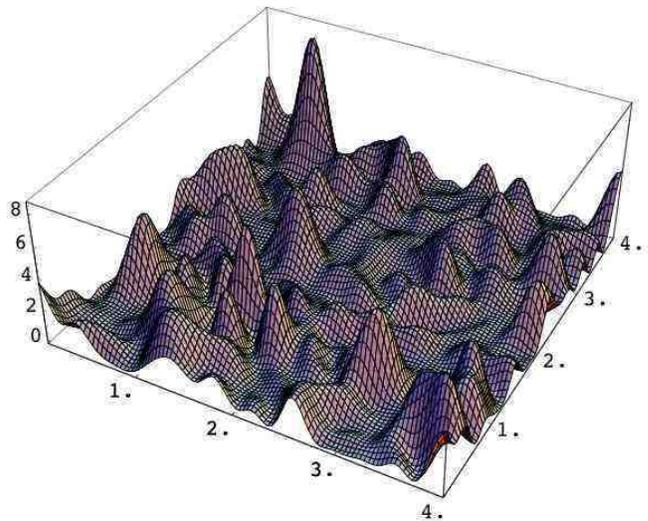}
\caption{Density distribution after the phase transition, at $\eta =
  10$ for the axion-like potential.  Axes are same as in
  fig.~\ref{white-noise}.  Note the highly nonlinear nature of the
  initial density perturbations.  This distribution remains fixed until near matter-radiation equality when it evolves gravitationally into collapsed ScaMs; this density spectrum is the input for the N-body simulation.}

\label{axion-like-initial}
\end{figure}

We evolve these fluctuations to $\eta = 10$, at which point the
fluctuations are expected to remain mostly spatially frozen (modulo logarithmic growth of fluctuations) until
gravitational collapse begins right around (or even somewhat before for the most dense ScaMs) the epoch of
matter-radiation equality.  The density fluctuations in a box with sides whose length are four times the horizon size at the time of the phase transition are shown in fig.~\ref{axion-like-initial}.
The density fluctuations have been normalized to the average density in the box, so that $\rho({\bf x})/\bar{\rho}$ is shown.   

We choose an alternate method to determine initial density
fluctuations for the Higgs-like potential.
According to eqn.~\ref{Higgs-like-flucs}, the Higgs-like potential gives only
quasi-nonlinear density perturbations, which allows us to model the formation of these clumps realizing the initial density fluctuations using  the standard N-body particle method: particles are initially placed on a lattice to minimize shot noise, then displaced from those positions by adding perturbations, mode by mode in the
Zel'dovich approximation, with amplitudes and phases selected
according to the distribution derived from the power spectrum.  We
adopt the power spectrum of eqn.~\ref{Pscam}
\be
P(k) = A e^{-(k r_s/2\pi)^2/2} \mbox{ Mpc}^3,
\ee
which creates white noise filtered on a scale $r_s$.  This smoothing scale
corresponds roughly to the horizon size at the time of the phase transition.  The
expected density fluctuations on this scale are thus
\be
\left(\frac{\delta \rho}{\rho}\right)_{rms}^2=\frac{9}{2\pi^2}k_s^2\int_0^\infty
P(k) \left(\frac{\sin(k/k_s)}{(k/k_s)^2}-\frac{\cos(k/k_s)}{k
    k_s}\right)^2 dk, 
\ee
where $k_s=2\pi/r_s$ and we choose $r_s$ so that the resulting ScaMs haves masses around
$1
M_\odot$ (calculated from $M_{ScaM} \sim \rho_{DM} r_s^3$, and corresponding to a phase transition temperature, $T_{trans} \sim 10^{-4} \mbox{ GeV}$), and $A$ such
that $\left(\delta \rho/\rho\right)_{rms} \simeq 0.5$.  Note that while we use specific physical
scales for the purpose of the simulation, the result is expected to be
completely scale invariant, and should apply with suitable rescaling to ScaMs of all
masses.

\begin{figure}
\includegraphics[width=8.5cm]{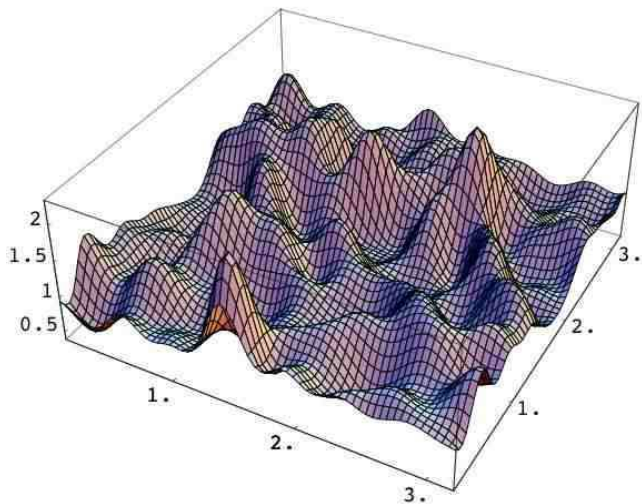}
\caption{Density distribution generated by the smoothed white noise power
  spectrum. The size of the density fluctuations were chosen
  consistent with the analytic result for the Higgs-like potential of
  Sec. II.  Like fig.~\ref{axion-like-initial}, this distribution is input for the N-body simulation.}
\label{slice}
\end{figure}

We plot the corresponding density fluctuations for the Higgs-like
potential in fig.~\ref{slice}.   We then use the N-body code to evolve the objects into the collapse epoch
near matter-radiation equality.

\subsection{N-body simulation of ScaM density profiles}

Using the density profile generated by solving the classical equations
of motion for the axion-like potential, a grid of $100^3$ particles is
laid with masses weighted according to the locally computed initial density.  The evolution of the
particles is started at a redshift $1\times 10^5$; objects
are fully collapsed by matter radiation equality, around a redshift of 3000.  As the
density fluctuations are initially much smaller for the Higgs-like distribution
shown in fig.~\ref{slice}, the evolution can be started much later; we
choose a redshift of $1 \times 10^4$, and objects are fully collapsed by a
redshift of $1000$.

\begin{figure}
\includegraphics[width=6cm]{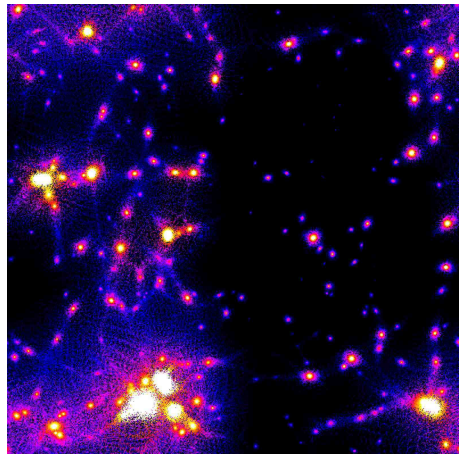}
\caption{Snapshot of structure formation at $z=3000$ for axion-like
  ScaMs; the plot is colored according to the logarithm of the density.   The scale here
  is $1.2 h^{-1}\mbox{ Mpc}$, in comoving coordinates, although the
  result is expected to be invariant for any box size, given that we
  rescale the smoothing scale accordingly.
  Note that structure formation has
  already commenced before matter radiation equality.}
\label{axion-like-snapshot}
\end{figure}

\begin{figure}
\includegraphics[width=6cm]{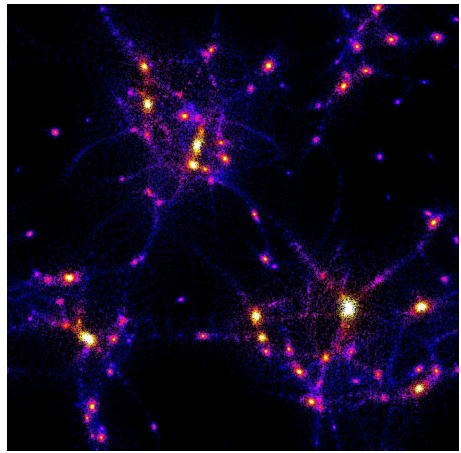}
\caption{Same as fig.~\ref{axion-like-snapshot}, but for the
  Higgs-like potential at a redshift $z=1000$.  This simulation was
  evolved to a lower redshift, as the initial over-density of the
  Higgs-like system is lower, and hence gravitational collapse occurs
  later.  Although difficult to see from this rendering, the densities
of the ScaMs are also lower.}
\label{Higgs-like-snapshot}
\end{figure}

We show in figs.~\ref{axion-like-snapshot},~\ref{Higgs-like-snapshot} a slice of the final particle distributions of the
collapsed objects, at redshifts around matter radiation equality.  The
initial over-densities shown in figs.~\ref{axion-like-initial},~\ref{slice}
evolve into the fully collapsed objects shown in
figs.~\ref{axion-like-snapshot},~\ref{Higgs-like-snapshot}.  The
pictures of course resemble images of the low-redshift cosmic web of
dark matter,  but in this case represent very small-scale, nonlinear
objects at a redshift around $10^3$. The ``normal'' inflationary perturbations in dark matter at this time still have a very small amplitude on all scales, of the order of one percent.

\begin{figure}
\includegraphics[width=8cm]{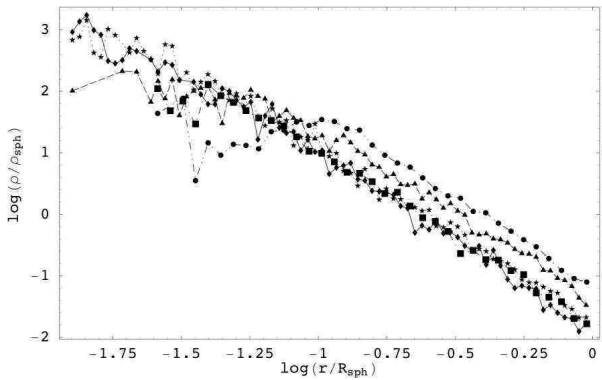}
\caption{Axion-like ScaM density profiles for five gravitationally
  collapsed ScaMs.  Vertical axis is $\log (\rho(r)/\rho_{sph})$,
  where $\rho_{sph}(\delta=1) = 280 \rho_{eq}$; that is, we normalized
  the density profile against the uniform density prediction of the
  spherical model, eqn.~\ref{isothermal-density}, with $\delta = 1$.  Horizontal axis is $\log (r/R_{sph})$, where $R_{sph}$ is the radius computed from $\rho_{sph}$ and the total mass contained within the horizon at the phase transition, $d_H(T_{trans})$.}
\label{axion-like-density}
\end{figure}

\begin{figure}
\includegraphics[width=8cm]{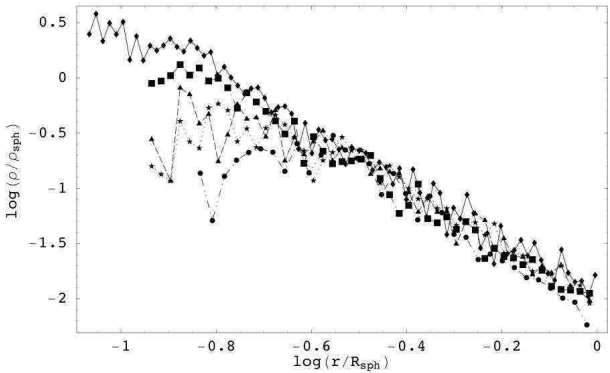}
\caption{Same as fig.~\ref{axion-like-density}, but for the Higgs-like
  potential.}
\label{Higgs-like-density}
\end{figure}

We plot in figs.~\ref{axion-like-density},~\ref{Higgs-like-density} the density profiles of a sampling of
ScaMs, where we choose to normalize our densities and radii
against those predicted by the spherical model,
eqn.~\ref{isothermal-density}, for $\delta=1$.  We
can see that the naive spherical model is approximately
accurate in predicting the maximum density of the ScaMs.  The axion-like potential has typical overdensities $\delta
\sim 5-10$, as shown in fig.~\ref{axion-like-initial}, so that the spherical model predicts $\rho/\rho_{sph}(\delta=1)
\sim 100 - 5000$.    The Higgs-like
potential, on the other hand, has typical overdensities $\delta \sim 0.5
- 1$ (fig.~\ref{slice}), and so we expect $\rho/\rho_{sph}(\delta=1) \sim 0.1 - 1$, again
consistent with the central densities given by the simulation.  The ScaMs with higher central densities correspond to ScaMs which had higher initial overdensity $\delta$.

%For
%the Higgs-like potential, we
%can see that the cores are denser than that predicted by the
%isothermal sphere approximation by a factor of
%$10-100$.  However, this is not enough to make these ScaMs
%visible for the current generation of  galactic scale lensing
%experiments.

\subsection{Strong lensing profiles}

What is the implication for lensing experiments?  As explained in section III, in order for an object to act as a strong gravitational lens, the enclosed mass, $M_{encl}$, at any given radius must exceed a minimum,
\be
M_{encl}(r) > 1 M_\odot \left(\frac{r}{s(D)}\right),
\ee
where $s(D)$ is dependent on the base length for lensing, $D$ ($s \sim 3\times 10^{15}\mbox{ cm}$ for $D = 5 \mbox{ Mpc}$, and $s\sim 3\times 10^{14}\mbox{ cm}$ for cosmological scale lensing).  Equivalently, at any given distance from the center of the object $r$, the radius must not exceed the Einstein radius,
\be
r < R_E = s(D) \left(\frac{M_{encl}}{1 M_\odot}\right)^{1/2}.
\ee
The Einstein radius can be computed directly from density profiles shown in figs.~\ref{axion-like-density},~\ref{Higgs-like-density}.  

\begin{figure}
\includegraphics[width=8cm]{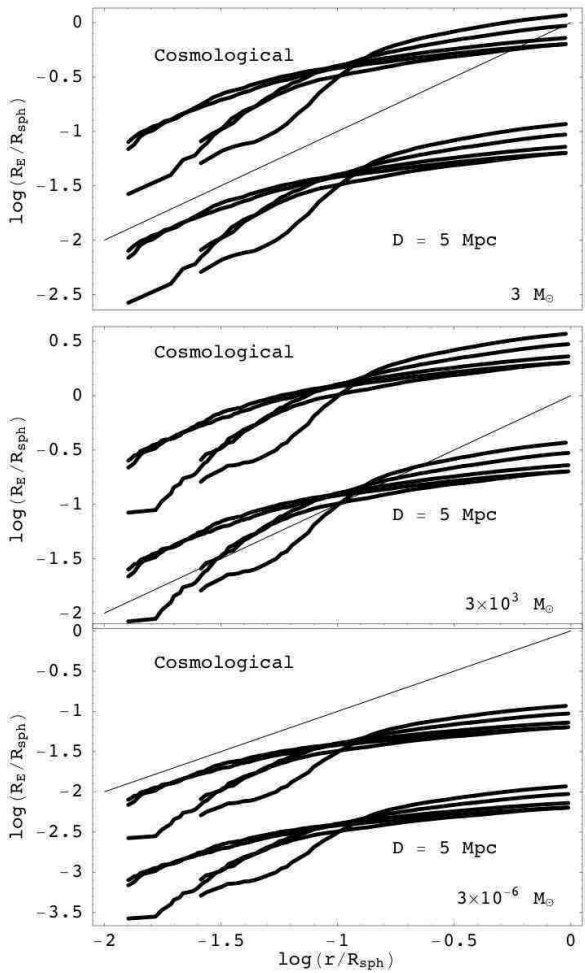}
\caption{Axion-like ScaM Einstein radius for the enclosed mass, derived from the computed N-body density profile, versus ScaM radius, again
  normalized against the ScaM radius $R_{sph}$ predicted by the
  spherical model. The diagonal line divides where $R_E > r$, when lensing
  is possible.  The upper set of curves in each plot is the Einstein radius for
  cosmological scale lensing, the lower set of curves the Einstein
  radius for lensing at a distance $D = 5 \mbox{ Mpc}$, which may be
  relevant for MACHO-type experiments in nearby galaxy halos.    This
  is shown for three different ScaM masses, marked at the lower right
  in each plot.}
\label{axion-like-Einstein}
\end{figure}

\begin{figure}
\includegraphics[width=8cm]{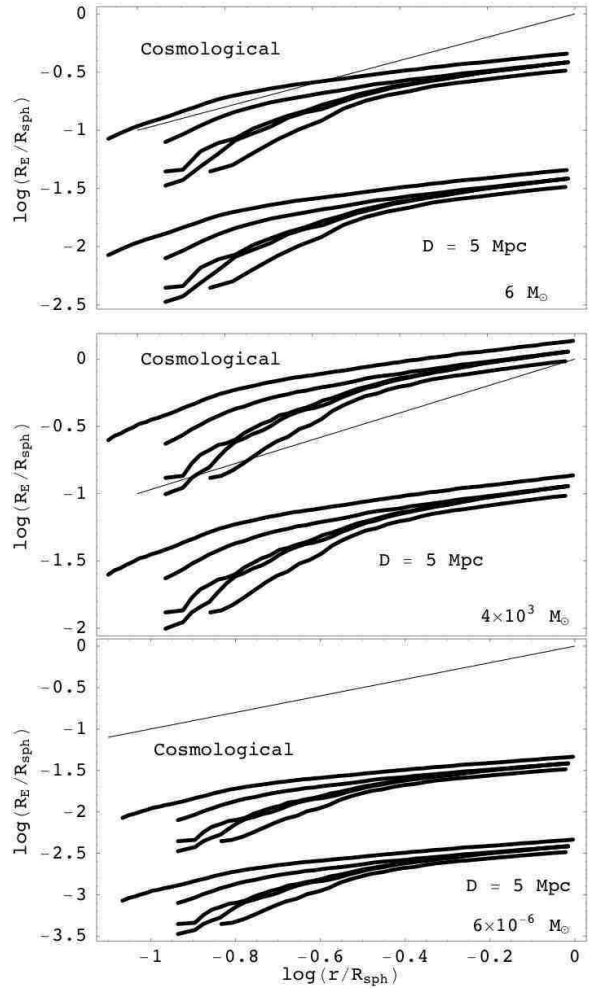}
\caption{Same as fig.~\ref{axion-like-Einstein}, but for the
  Higgs-like potential.}
\label{Higgs-like-Einstein}
\end{figure}

We plot in figs.~\ref{axion-like-Einstein},~\ref{Higgs-like-Einstein} $R_E$
versus $r$, for $s=3\times 10^{15}\mbox{ cm },~3\times 10^{14}\mbox{
  cm}$.  As in figs.\ref{axion-like-density},~\ref{Higgs-like-density}, we have normalized the radii against
$R_{sph}$, the prediction of the spherical model.  The objects lens
if $R_E > r$, that is, if the Einstein profiles lie above the straight
line, $R_E = r$, shown in the figure.  Although the simulation was run
for a particular mass within the horizon at the time of the phase
transition, $M \sim 1 M_\odot$, it is simple to apply the results to
many different ScaM masses simply by rescaling the total mass in the
box.  In this scaling $R_E$
increases like $M^{1/2}$, but the box size decreases like $M^{1/3}$ so
the net result is that the vertical axis in the figure is scaled up by
$M^{1/6}$, meaning that more massive ScaMs lens more easily.
Note that the mass labeled is the mass enclosed within the horizon at
the time of the phase transition, not the lensed mass of the ScaM, which may be significantly lower.

We can see that both Higgs-like and axion-like ScaMs are of
interest for lensing experiments on a cosmological scale, from
objects as light as $10^{-6} M_\odot$ to as heavy as the cosmological
bound of $4\times 10^3 M_\odot$.  While neither of these objects is of
interest for the current galactic lensing experiments with baseline $D
\sim 50 \mbox{ kpc}$, the axion-like configuration in particular will
be of interest for lensing  in the Mpc range and greater, as might be accessible in future experiments.   We also note here that we have not chosen
the most extreme set of initial conditions to simulate for the
axion-like potential.  The axion itself has a time dependent mass
around the time of the phase transition which results in larger
density contrast.  Normalizing against the size of the box, we find
that axions in particular generate overdensities $\delta \sim 20-30$. (Kolb
and Tkachev \cite{Kolb:1994fi} found larger overdensities $\delta \sim 30-100$ as they
normalized against the density for $\theta_{rms} = \pi/\sqrt{3}$ and
not the actual average density in the box.)  In this case, $R_E$ for
the axion-like potential is increased by as much as a factor of 10,
which improves the observational prospects.

As long as the radius of the lens is much bigger than the size of the
background source, those situations where $R_E>r$ will occasionally
lead to strong-lensing events with high amplification. The frequency
of this happening is roughly given by the fraction of  the halo mass
in ScaMs above this threshold, times the mean lensing optical depth of
the halo, times $A^{-2}$ where $A$ is the amplification. The mean
optical depth is a very small for nearby halos ($\approx 10^{-6}$ for
the LMC experiments) but  is typically $\approx 0.1$  for  halos at
cosmological distances. Such considerations affect the number of
sources and the cadence of observations needed in a variability
survey.

\subsection{Weak lensing: mapping ScaMs using variability}

Even if    ScaMs are not dense enough to cause strong-lensing events, 
they can in general still produce   weak lens amplification.   In 
some situations   the
systems are small enough that this amplification is time variable, 
and  the
projected density profiles of ScaMs can be measured directly by 
monitoring
sources. Survey parameters, such as the distance of the halo under 
study and the
number of sources to be monitored, can be optimized depending on the 
predicted
ScaM parameters; in principle, both  weak and strong lensing can be 
studied at
various distances.

\begin{figure}
\includegraphics[width=8cm]{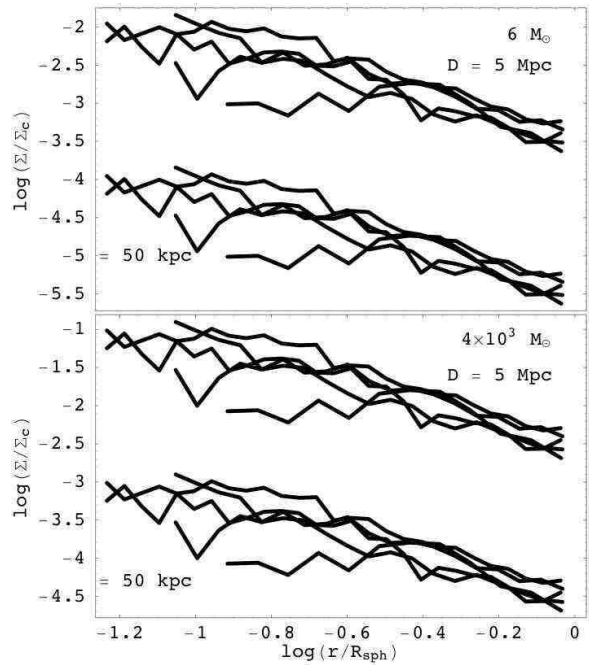}
\caption{Log mean surface density $\Sigma$ relative to the critical 
density
   $\Sigma_c$ as a function of log radius, averaged in annuli, 
normalized again to the
   prediction of the spherical model, $R_{sph}$.  $\Sigma/\Sigma_c 
\sim 1$
   corresponds roughly to the onset of strong lensing, and
   $2 \Sigma/\Sigma_c$ corresponds to the fractional weak-lens
   amplification.  Higgs-like potential. }
\label{Bias4e7SurfaceDensity}
\end{figure}

\begin{figure}
\includegraphics[width=8cm]{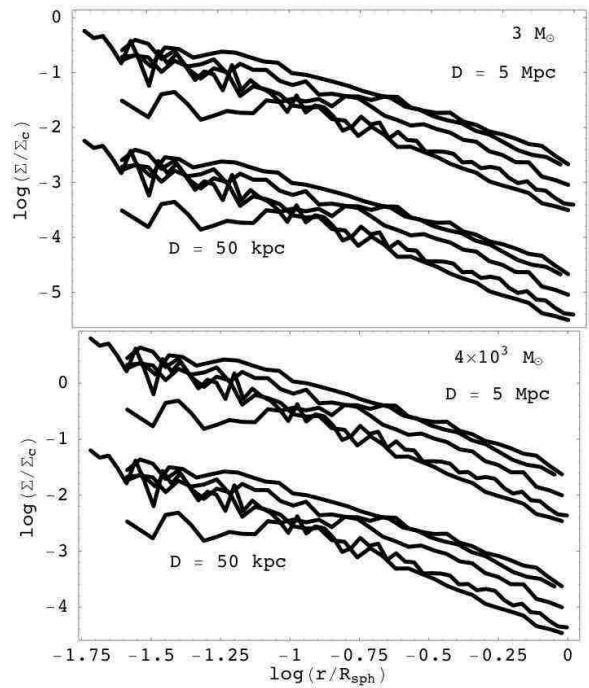}
\caption{Same as fig.~\ref{Bias4e7SurfaceDensity}, but for the
   axion-like potential.}
\label{SineSurfaceDensity}
\end{figure}

Consider a standard gravitational lens mapping a source at position $
\vec
\theta_S$ in the source plane to a position $\vec \theta_I$ in the 
image plane.
In the image plane the mapping is characterized by the convergence $
\kappa$ and complex shear
$\gamma$.
The magnification $\mu$ is the inverse Jacobian of  the mapping \cite{Bartelmann:1999yn}
\be
\mu^{-1}= (1-\kappa)^2-|\gamma|^2.
\ee
The convergence is determined by the surface density $\Sigma$ along 
the line of
sight,
\be
\kappa=\Sigma/\Sigma_C,
\ee
where the critical surface density depends on the (angular diameter) 
distances to
the source ($D_S$), the lens ($D_L$), and between them ($D_{LS}$) as:
\be
\Sigma_C=D_{LS}c^2/4\pi G D_LD_S\approx 3.   5 {\rm kg\ m^{-2}}(D_{LS}
{\rm \
Gpc}/D_LD_S).
\ee
If $\Sigma>\Sigma_C$, generally light rays focus to a point 
somewhere, the mapping involves
multiple images and generally strong shear, and amplification is both 
nonlocal and nonlinear. High-amplitude variability is in the strong 
lensing regime and dominated by caustics near fold catastrophes in 
the mapping.
By contrast, in the case of weak lensing, $\Sigma<\Sigma_C$, the 
amplification of a source is
more directly related to the surface density in that direction. For 
small surface densities, and where the shear is negligible, the 
magnification is just
\be
\mu=\left |{\partial \vec \theta_S\over \partial \vec \theta_I}\right 
|^{-1}\simeq 1+2\kappa.
\ee
Thus as a source traverses behind a ScaM and $\Sigma$ varies, its 
brightness
changes by a fraction $2\Sigma/\Sigma_C$.  We plot in
figs.~\ref{Bias4e7SurfaceDensity},~\ref{SineSurfaceDensity}
$\Sigma/\Sigma_C$ for the axion-like and Higgs-like potentials.  We
see that the variability will be difficult to detect for the current
lensing experiments towards the LMC and M31, as the effect is only a few 
percent even for the most massive ScaMs.   The weak lensing 
variability, however, will be of interest for longer
baseline experiments, $D = 5\mbox{ Mpc}$, and also for cosmological 
scale
lensing.  For an axion-like ScaM of mass $1 M_\odot$,
the time scale for variability is on the order of a year (assuming a
ScaM velocity across the source of 300 km/s), and decreases with ScaM 
mass as $M_{ScaM}^{1/2}$.

Note that the weak lens effects cover more area, and have a larger 
probability of
affecting a background source than the strong lens effects. For  a 
given mass of lensing material, the
lensing probability or optical depth scales in the weak regime  like
$\tau\propto \Sigma^{-1}\propto (\mu-1)^{-1}$, but the distribution 
of mass in the outer parts of ScaMs is even more favorable to the 
weak lensing program. In the examples shown in figures  11 and 12, we 
see that variability at the ten percent level
(that is, $\Sigma/\Sigma_C\approx 0.05$) occurs at a radius which is 
typically more than ten times the radius of strong lensing, so 
variation of this magnitude is over 100 times more frequent than 
strong lensing events.   Photometric accuracy and stability are the 
main practical limits in mounting variability surveys around this 
effect, but long term variations at the few percent level are 
certainly within the range of proposed instruments (such as JDEM) 
designed to monitor distant supernovae.

%These
%comparatively rare objects with very large overdensities would in fact
%be visible by galactic scale lensing experiments if their masses were
%in the range $10^{-7} M_\odot - 10 M_\odot$ currently accessible.

%\section{Partial Lensing}

%These objects are extended with radius given by eqn.~\ref{R} which is within about an order of magnitude of the einstein radius.  This gives rise to the possibility of partial lensing by the ScaMs, occuring when the rays from the source pass directly through the lens.  The probability for this to occur when the source is outside the galaxy is
%\be
%P = \pi R_c^2 \int_{R_0}^{R_{H}} n_c r^2 dr,
%\ee
%where $R_0$ is the sun's distance from the center of the galaxy, about $8 \mbox{ kpc}$, $R_H$ is the radius of the dark matter halo, taken to be $\sim 40\mbox{ kpc}$, and $n_c$ is the number density of ScaMs, derived from the NFW density profile
%\be
%\rho_{DM} = \frac{\rho_0}{r/r_s(1+(r/r_s)^2},
%\ee
%where $\rho_0 \simeq 5 \times 10^{-2} M_\odot/\mbox{pc}^3$, and $r_s \simeq 5 \mbox{ kpc}$.  Then
%\be
%P = 2 \times 10^{-5} \frac{1 M_\odot}{M_c},
%\ee
%so that the prospects to measure such partial lensing are small without a large number of events.  Prospects are improved by an order of magnitude measuring lensing toward the galactic center, but still about 5,000 events would be necessary to get a single partially lensed event for a $1 M_\odot$ object.

\section{Early baryon collapse}

With WIMP-type dark matter such as neutralinos, inflationary fluctuations lead to collapse of the first dark matter halos of roughly an earth mass scale, determined by the damping scale of the WIMP streaming motions \cite{Diemand:2005vz}. The collapse begins on after a redshift of 100 and on small scales has little effect on the baryons, since the gravitational potentials are so shallow. 

With ScaMs on the other hand, the addition of isocurvature fluctuations creates deeper dark matter potentials at earlier times. We have already seen that the accretion of absorbing gas, leading to Ly$\alpha$ absorption, is indeed the main astrophysical constraint at present on ScaM mass. 
In this section, we estimate the nonlinear effects of ScaMs on early structure,   using an analytic Press-Schechter approach to bottom-up hierarchical clustering.

Once the ScaMs form at $z_{eq}\approx 3300$, bottom-up hierarchical structure formation begins.  In the spherical model, fluctuations on a scale $M$ collapse into structures once the fluctuation has grown to a size
\be
\delta(M) \simeq 1.69.
\ee
As the fluctuations are isocurvature, they grow as
\be
\delta(M) = \delta(M_{ScaM}) \sqrt{\frac{M_{ScaM}}{M}} \frac{3}{2}\frac{1+z_{eq}}{1+z},
\ee 
where $\delta(M_{ScaM}) \simeq 1$.  Therefore a structure of mass $M$ will form at a redshift $z_f$
\be
1+z_f \simeq \frac{1}{\delta(M)}\sqrt{\frac{M_{ScaM}}{M}} \frac{3}{2}(1+z_{eq}).
\ee
By making use of the spherical approximation, $\rho_M = 140 \delta(M)_i^3(\delta(M)_i+1) \rho_{eq}$, where
\be
\delta(M)_i = \delta(M_{ScaM})  \sqrt{\frac{M_{ScaM}}{M}}
\ee
we also determine the virial radius of a structure of mass M,
\be
R(M) \simeq \left(\frac{M}{6\times10^2 \delta(M_{ScaM}) \rho_{eq}}\right)^{1/3} \sqrt{\frac{M}{M_{ScaM}}},
\ee
and hence the virial velocity of these structures is
\beq
\sqrt{GM/R} & \simeq & G^{1/2}\left(6\times10^2 \delta(M_{ScaM}) \rho_{eq}\right)^{1/6} (M M_{ScaM}^3)^{1/12} \nonumber \\
        & \simeq & 5 \times 10^{-6}\left(\frac{M}{1 M_\odot}\right)^{1/12}\left(\frac{M_{ScaM}}{1 M_\odot}\right)^{1/4}.
\eeq

If the baryons have a thermal velocity greater than the virial velocity, they are prevented by pressure gradients from collapsing into the ScaM.  To trigger accretion and collapse we require the ScaM to have a virial temperature greater than that of the microwave background, since the baryons can certainly not radiate binding energy if they are cooler than that.
The velocity of the baryons at the CMB temperature today would be $v \simeq \sqrt{3 T_0/m_p} \simeq 9 \times 10^{-7}$.  The baryons begin accretion onto the ScaM seeded structures once the velocity of the baryons drops below the structures' virial velocity, which occurs at redshift, $z_{acc}$,
\beq
2 \times 10^{-6}&\left(\frac{M_{ScaM}}{1 M_\odot}\right)^{1/3}\left(\frac{3}{2\delta(M)}\frac{(1+z_{eq})}{(1+z_{acc})}\right)^{1/6} \\ \nonumber 
&\simeq
\sqrt{\frac{3 T_0 (1+z_{acc})}{m_p}}.
\eeq
Solving this, we find that baryon accretion onto the structures formed hierarchically from a ScaM seed of mass $M_{ScaM}$, occurs at a redshift
\be
1+z_{acc} \simeq 30 \left(\frac{M_{ScaM}}{1 M_\odot}\right)^{1/2},
\ee
onto a structure of mass
\be
M = \frac{1 M_\odot}{\delta(M)^2} \left(\frac{1+z_{eq}}{1+z_{acc}}\right)^2.
\ee

We do not pursue further here the physics of the accreted baryons,
which depends on complex details of nonlinear collapse and cooling
\cite{Ripamonti:2005ri,Abel:2001pr,Yoshida:2003rw,2001PhR...349..125B}.
However it is clear that the early star formation in the ScaM models can start much earlier than in standard CDM cosmology where significant collapses occur at a redshift less than about 30. With massive ScaMs, collapse can occur as early as recombination. Since cooling is relatively efficient at such early times (for example, through Compton cooling on the microwave background), these models also likely produce very early stars.  These effects may eventually be observable either directly via deep infrared imaging, or indirectly via reionization effects. In any case it is clear that there would be significant modifications to early star formation with standard CDM perturbations.

\section{Conclusion}
We have not attempted to trace the evolution of the ScaMs in detail to
the present, where they would be incorporated hierarchically into the
standard galaxy-size dark matter halos. Although they would be subject
to some disruption in the course of hierarchical assembly, there is
good reason to think that they would mostly  survive intact to the
present. Certainly they fare better than the much more diffuse
earth-mass halos of neutralino CDM \cite{Diemand:2005vz}. In that case,  the very flat distribution of density leads to a large range of masses collapsing almost at the same time on top of each other.   Early ScaMs cluster instead with a steep white-noise spectrum, where the mass scale grows as a power law. In that situation there is more room for survival and   less chance for disruption; the process of early ScaM clustering has a similar spectrum to, and therefore resembles, the  larger galaxy-scale hierarchy today, where simulations have established that much of the satellite substructure survives.  
The clumpiness of dark matter in the halo may also have observable
consequences through tidal forces, such as  disruption of globular
clusters; these effects have been studied in the context of black hole
dark matter and other highly compact objects \cite{Arras:1998ib,Moore:1993sv,2004ApJ...601..311Y}. 

We have shown that in scalar theories with a cosmological phase
transition below the QCD scale, ScaMs form with a mass in an interesting range for microlensing experiments, $10^{-12}
M_\odot \lesssim M_{ScaM} \lesssim 10^4 M_\odot$.  The radii and density profiles of the
ScaMs vary considerably, depending on the initial size of the
density fluctuations.  For an overdensity $\delta \sim 50$ and mass $M
\sim 1 M_\odot$, these objects would be visible for current
galactic scale lensing experiments.  Objects with (more plausible) lower overdensities
$\delta \sim 5-10$, while not detectable with the current generation of experiments, would be visible for future lensing experiments with
longer baselines, $D \sim 5 \mbox{ Mpc}$ or greater, requiring   a space based platform (or wide field adaptive optics) to conduct a lens-induced-variability survey.   Cosmological scale lensing
is the most powerful tool of all for detecting these objects, but on that scale,  detailed monitoring studies are still in the distant future.  Some information might be obtained sooner from quasar microlensing, but ScaMs tend to be  not much larger than quasar emitting regions, so this technique has limited application and dynamic range. Smaller sources such as individual stars are, of course, much fainter.

The scalar field composing these objects is so weakly coupled to the standard model that they would never be detected directly, but would only make their effect known to us gravitationally. Thus the rich substructure of the halo dark matter does not have other effects, such as gamma ray annihilation signatures or clustering in direct detection experiments, that occur for other dark matter candidates. The main exception to note is the possibility of ScaMs of classical invisible axions, which could show up in direct detection as clumping in time as well as energy.

ScaM particles when they condense are moving with modestly relativistic velocities in spatially coherent streaming flows on the scale of the nonlinear lumps at that time. These redshift by the time of the nonlinear collapse so that  theydo not prevent collapse, but it is worth mentioning that they do not have the 
dynamically-cold classical distribution function characteristic of homogeneous axion condensation.  The broader coarse grained distribution function could in principle  have some dynamical effects, which are too subtle to model in our simulations.

The scalar field collapses into ScaMs at matter-radiation
equality, seeding early bottom up hierarchical structure formation as
successively larger mass scales become virialized. This in turn results
in early star formation, with baryon accretion onto the Scams
starting as early at $z \sim 1400$ for a ScaM mass near the
cosmological limit at $M_{ScaM}\sim 4\times 10^{3} M_\odot$.  This could have substantial observational effects on energy input into diffuse gas, affecting  the epoch of reionization, and possibly also direct detection of early stellar systems. The reionization epoch is already being probed by quasar absorption to $z>6$ and by CMB polarization to $z>10$; even larger redshifts may become accessible in the future to similar techniques, as well as direct 21cm mapping. These more indirect effects are of course more complicated to model than the more direct influence of the lensing.

The nature of the dark matter remains unknown to us.  Scalar fields
play an important role in cosmology, being front and center in viable
models of inflation and many theories of the dark energy.  We have
shown here that, should a scalar field also contribute to solving the
dark matter mystery, it is a good candidate for large isocurvature
fluctuations which cause them to collapse into dense ScaMs; in fact,
though the scalar may live in a hidden sector extremely weakly coupled
to us, so that we would detect no direct particle interactions,
gravitational lensing may provide us a probe into this sector. 

\begin{acknowledgements}
This work was supported by NSF grant AST-0098557 at the University of
Washington.
\end{acknowledgements}

\appendix*
\section{Late Phase Transitions in Axion Cosmology}

We demonstrate a technically natural model which provides a phase transition well below
the QCD scale, seeding density fluctuations in both Higgs-like and
axion-like fields.  This model was motivated in
ref.~\cite{Kaplan:2005wd} for the purpose of demonstrating that in
theories where the axion's decay constant evolves after the QCD phase
transition, the cosmological upper bound on the axion decay constant,
$f_a \lesssim 10^{12} \GeV$, does not apply. The model as written here
applies specifically to the QCD axion, but the model can be
generalized to any set of scalar fields, not necessarily connected
with the solution of the strong CP problem.

%For the purposes of this
%paper, we decouple the model from the implications for the QCD axion
%to demonstrate the more general class of models relevant for the
%ScaM physics.

%The late phase transition, below the QCD scale, was introduced in
%ref.~\cite{Kaplan:2005wd} for the purpose of demonstrating that in
%theories where the axion's decay constant evolves after the QCD phase
%transition, the cosmological upper bound on the axion decay constant,
%$f_a \lesssim 10^{12} \GeV$, does not apply.  

The phase transition in this model is driven by a complex field
$\phi$, whose radial mode has a Higgs-like potential,
\be
V(\phi) = \mu^4\left|\frac{\phi^2}{M^2}-1\right|^2,
\ee
where we require $\mu^2/M \ll H(T_{QCD})$ in order that the vacuum
expectation value (vev) of the field
remain frozen at its initial position, away from the minimum of the
potential at $\langle \phi \rangle =  M$, until $T < T_{QCD}$.  The
field $\phi$ is coupled to the Peccei-Quinn (PQ) sector through a potential
\be
V(\phi,X_1,X_2)=\lambda^2|hX_1X_2-\phi^2|^2,
\label{flat}
\ee
where $X_1$ and $X_2$ carry opposite PQ charges.  $X_1$ is the
standard PQ field coupled to gluon field strength through the term
\be
X_1G\tilde{G},
\ee
where the usual PQ potential fixes the vev
$|X_1|=f_1$ when the PQ symmetry breaks at temperatures $T \gg
T_{QCD}$.  At the QCD phase transition, the axion gains a mass $m_a \sim
m_\pi f_\pi/f_1$, and a dark matter condensation of axions forms.

A phase transition occurs when $m_\phi \sim H(T)$, so that $\phi$
rolls out to the minimum of its potential at $|\phi|=M$; $|X_2|$ then follows
the flat direction in eqn.~\ref{flat} to a vev $f_2 = M^2/h f_1$,
where the hierarchy $f_2 \gg M \gg f_1$ is assumed.  It can then be
shown (see \cite{Kaplan:2005wd} for details) that the energy density in the axion dark matter is transferred into
the heavier pseudo-scalar $\pi_1 \equiv f_1 Arg(X_1)$, having mass
$m_{\pi_1} ~ \lambda M^2/f_1$ generated at the phase transition.

As a result of the phase transition, there are two dark matter candidates in this model, one Higgs-like and
one axion-like.  At the phase
transition, the Higgs-like mode $|\phi|$ releases energy $\mu^4$ which subsequently
redshifts like dark matter, and may be cosmologically abundant.  It is
so weakly coupled to ordinary matter that it may only be detected
gravitationally.  It has ${\cal O}(1)$ density perturbations resulting
from the phase transition at a temperature $T \ll T_{QCD}$, which
collapse into Higgs-like ScaMs with mass much larger than an axion
minicluster, $M_{ScaM} \gg 10^{-12} M_\odot$.  The axion-like
pseudoscalar $\pi_1 $ may also be cosmologically
populated, and collapses into axion-like ScaMs.  The mass of
these ScaMs is much larger than QCD axion miniclusters
as the pseudo-Goldstone boson in this model has an additional mass
generated at the phase transition $m_{\pi_1}/m_a 10^{-12} M_\odot \gg 10^{-12} M_\odot$.  

There is a mechanism in this model which allows for equal cosmological
abundances of Higgs-like and axion-like dark matter, $\phi$ and $\pi_1$, without any fine-tuning.  At the QCD phase transition, $\pi_1$ dark matter is produced with energy density
\be
\rho_{\pi_1}(T) = m_{\pi_1} n_{\pi_1},
\ee
where $n_{\pi_1}$ is the number density, which dilutes as $T^3$, and $m_{\pi_1}$ now receives a contribution from its mixing with $\phi$:
\be
m_{\pi_1}^2(\phi) \sim \frac{m_\pi^2 f_\pi^2}{f_1^2}+\frac{\lambda^2
  \langle |\phi|\rangle^4}{f_1^2},
\ee
This creates an effective potential for $\phi$,
\be
V_{eff}(\phi) = m_{\pi_1}(\phi) n_{\pi_1}.
\ee 
This effective potential will delay the temperature of the phase
transition, when $\phi$ rolls to $\langle |\phi|\rangle = M$, until $\pi_1$ has
diluted enough that $V_{eff}(M) < \mu^4$.  At that point the energy
densities in the $\phi$ and $\pi_1$ fields are equal, and they
redshift concomitantly so that their energy densities remain equal
thereafter.  This gives rise to the possibility of the cosmological
presence of both Higgs-like and axion-like miniclusters. As the
Higgs-like and axion-like miniclusters have different typical
densities (the latter often being $10^4$ times more dense, as
explained in Sec.~III), this implies a potentially varied cosmic
population reachable by lensing experiments on different scales.

%For the purposes of this paper, we will focus on the dark matter candidate $\phi$ which drives the phase transition;  we will show that for phase transition temperatures, $T_{trans}$, which are of interest for this model, dark matter ScaMs form with masses and radii which make them detectable by microlensing experiments.  We will show that these scalar miniclusters may in fact explain the excess of objects observed by microlensing experiments MACHO and POINT-AGAPE in the mass range $M_{ScaM} \sim 0.1 M_\odot$.

We also wish to emphasize that the physical features discussed here of
late scalar field phase transitions are quite generic and independent
of the presence of axions; however, the axion model provides a
motivation to consider such phase transitions around the QCD time, as
well as illustrates a model with all the physical features discussed
in this paper.

\bibliography{latephase.bib}

\end{document}